# Topographic patterning in perovskite oxide membranes for local control of strain, nanomechanics and electronic structure

Marti Ramis[1], Markos Paradinas[1], Jose M. Caicedo[2], Claudio Cazorla[3,4,5], Roger Guzman[1], Mariona Coll[1,*]

[1]Institut de Ciència de Materials de Barcelona (ICMAB-CSIC), Campus UAB, 08193 Bellaterra, Barcelona, Spain
[2]Catalan Institute of Nanoscience and Nanotechnology (ICN2) Campus UAB, Bellaterra 08193, Barcelona, Spain
[3]Department of Physics, Universitat Politècnica de Catalunya, 08019 Barcelona, Spain
[4]Research Center in Multiscale Science and Engineering, Universitat Politècnica de Catalunya, 08019 Barcelona,S [5]Institució Catalana de Recerca i Estudis Avançats (ICREA), Passeig Lluís Companys 23, 08010 Barcelona,
Spain

mcoll@icmab.es

Single-crystalline perovskite oxide membranes provide a powerful platform to access physical properties that are inaccessible in bulk crystals and substrate-clamped thin films. Within this context, the deliberate fabrication of tailored corrugations provides a reliable mean to impose local curvature enabling deterministic modulation of functional properties. Here, we demonstrate controlled topographic patterning in (00l)-oriented $La_{0.7}Sr_{0.3}MnO_3$ (LSMO) membranes with thicknesses ranging from 4 to 100 nm where they spontaneously form sinusoidal wrinkles with thickness-dependent periodicity and amplitude. The wrinkle morphology directly modulates membrane stiffness and generates exceptionally large local strains exceeding 5% with strain gradients approaching ∼ 2.5 x $10^7$ $m^{-1}$ in the thinnest membranes. These extreme deformations suppress antiferrodistortive octahedral rotations and stabilize polar distortions, evidencing a curvature-driven symmetry transformation. The surface potential variation reinforces the formation of wrinkled-induced polar patterns being strongly modulated with thickness. The variation of Mn oxidation state from ∼ 3.2+ to ∼ 2.85+ provides a direct chemical signature of a thickness-controlled electronic transition. These results demonstrate that corrugation-induced strain gradients in oxide membranes with different thicknesses can drive coupled structural, nanomechanical and electronic transformations offering a singular route to engineer their functional states for next-generation electronic devices.

## 1 Introduction

The formation of topographic patterns such as wrinkles has driven intense research in the development of smart, stretchable surfaces with widespread applications in electronics, [1, 2, 3] energy [4] and biomedicine.[5, 6] For the classical case of a stiff film on a compliant substrate, spontaneous formation of wrinkles occurs when interfacial in-plane compressive stress, induced by mechanical or thermal mismatch, exceeds a critical strain threshold.[7, 8, 9, 10] In turn, controlled geometric design of the wrinkle, enabled by tuning the thickness and elastic modulus of the film, establishes a compelling strategy to locally engineer their mechanical and electronic properties. [11, 12, 13, 14, 15, 16, 17]

In strongly correlated oxides—where charge, spin, orbital, and lattice parameters are intimately coupled—, the precise design of wrinkle patterns offers a powerful avenue for expanding and tailoring their diverse functional landscape. [18] However, the intrinsic brittleness and mechanical rigidity of these materials in bulk form or as epitaxial thin films severely limited their ability to accommodate the large deformations required to access beyond epitaxial strain-engineered phenomena. The fabrication of freestanding complex oxide membranes[19] has transformed this constraint, establishing new paradigms in material design and heterointegration that allow unprecedented control over functional properties and provide access to emergent functionalities. [20, 21, 22, 23, 24, 25, 26, 27] Recent pioneering reports on spontaneous wrinkling in oxide membranes have revealed distinct strain-correlated responses generated by the flexoelectric effect.[28, 29] For example, stabilization of ordered polar structures in otherwise nonpolar SrTiO$_3$ (STO) [30] and marked enhancement of piezoelectric response in BaTiO$_3$ (BTO).[21]

Nevertheless, a reliable and facile approach for designing periodic wrinkles as well as a subsequent comprehensive nanoscale characterization is missing. This gap hinders a robust understanding of how wrinkle curvature couples to atomic and electronic structure, and how membrane thickness mediates this coupling thereby impeding direct morphology-structure-property relationship that is essential for the rational design of next-generation flexible and smart nanoelectronic devices.[31, 32, 33]

La$_{0.7}$Sr$_{0.3}$MnO$_3$ (LSMO), is a prototypical strongly correlated oxide that exhibits a rich phase diagram including colossal magnetoresistance, a metal-insulator transition, and a Curie temperature above room temperature. [34, 35, 36, 37] As such, LSMO is highly sensitive to lattice distortions[38, 39, 40] and is a particularly well-suited model system for investigating the spatial modulation of correlated electronic behaviour through controlled surface wrinkling. In this work, we fabricate a series of (00l)LSMO membranes with thickness ranging from 4 to 100 nm on a compliant silicone/polyethylene terephthalate (PET) elastomeric substrate using a water-soluble sacrificial layer strategy,[41] which enables tunable wrinkle formation by tailoring the film–substrate interaction. We combine multimodal morphological, structural, mechanical and spectroscopic probes to resolve how wrinkling reshapes the properties of LSMO membranes. Atomic force microscopy (AFM) and force distance spectroscopy (FD), are used to map the wrinkle topography and the local stiffness of the system. Scanning transmission electron microscopy (STEM) together with electron energy loss spectroscopy (EELS) quantify the atomic distortions, strain and electronic structure. Then, by means of electrostatic force microscopy (EFM) combined with Kelvin probe force microscopy (KPFM) we mapped the surface electric field distribution as a function of the membrane morphology and thickness. Together, this comprehensive study enables designing a rich thickness-dependence phase diagram where wrinkled-induced strain and strain gradient effects in LSMO membranes deliver local modifications of the stiffness and polar structure alongside a pronounced thickness-dependent conductivity.

## 2 Results and discussion

### 2.1 Crystallinity

A series of La$_{0.7}$Sr$_{0.3}$MnO$_3$ (LSMO) films with thicknesses ranging from 4 to 100 nm were grown on 20 nm-thick (00l)SrCa$_2$Al$_2$O$_6$ (SC$_2$AO) sacrificial layer deposited on SrTiO$_3$ (STO) substrates. X-ray diffraction (XRD) $\theta$-$2\theta$ scans of the resulting LSMO/SC$_2$AO//STO heterostructures (Figure 1a) confirm the (00l) oriented growth of both LSMO and SC$_2$AO on STO. The invariant position and intensity of the (008) SC$_2$AO Bragg reflection indicate that the sacrificial layer remains structurally intact during LSMO deposition. With increasing LSMO thickness, the (002) LSMO reflection, which appears as a shoulder in thinner films, becomes progressively more intense in thicker ones. Reciprocal space maps collected around the (103)STO reflection (Figure S1, Table S1) reveal that the SC$_2$AO lattice parameters remain constant with $c_{SC2AO}$ =3.84 Å and $a_{SC2AO}$=3.86 Å, close to the bulk pseudo-cubic lattice parameter, $c_{SC2AO-bulk}$ = 3.852 Å.[42] For LSMO, the lattice parameters could be unambiguously extracted for 50 and 100 nm films, yielding $a_{LSMO}$= 3.86 and $c_{LSMO}$ =3.88 Å, indicating that compared to the pseudocubic $c_{LSMObulk}$= 3.876 Å, the LSMO films experience a slight in-plane compressive strain (0.1%).

Following growth, the LSMO films were transferred onto silicone-coated polyethylene terephthalate (PET) substrates by bonding the elastomeric support to the film surface and selectively dissolving the SC$_2$AO layer in Milli-Q water. XRD measurements of the transferred LSMO/PET membranes reveal diffraction peaks corresponding to both the PET substrate and the (00l)-oriented LSMO (Figure 1b). The intensity of the LSMO reflections scales systematically with membrane thickness, as highlighted in the inset, consistent with the expected dependence. The in-plane ($\Delta\phi$) and out-of-plane ($\Delta\omega$) crystalline texture of the membranes was quantified for all thicknesses, with the corresponding values summarized in Table S2. Both $\Delta\phi$ and $\Delta\omega$ decrease with increasing membrane thickness, indicating a progressive improvement in crystalline texture, with the 100 nm membrane exhibiting the highest crystalline quality among the series. Nevertheless, these values remain modestly broader than those typically reported for lattice-matched LSMO films grown directly on STO substrates, which is attributed to the cumulative effects of multilayer heteroepitaxy and the membrane transfer process. [43]

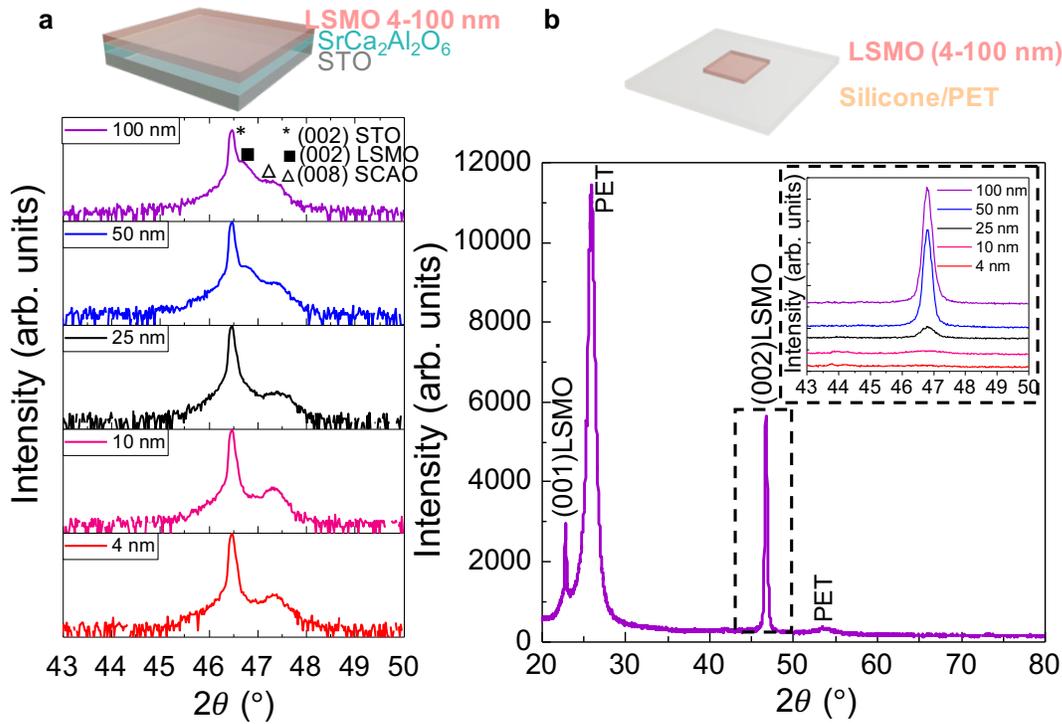

Figure 1: XRD $\theta$-$2\theta$ scans from (a) LSMO/SC$_2$AO//STO heterostructures with varying LSMO film thickness (4-100 nm). (b) 100 nm (00l)LSMO membrane on PET . Inset shows (002) LSMO reflection of the LSMO on PET for the different film thicknesses.

## 2.2 Wrinkle Morphology

The surface morphology of the LSMO membranes on the PET sheet was investigated using optical microscopy and atomic force microscopy (AFM), as shown in Figure 2. Optical micrographs (Figure 2a) reveal the spontaneous formation of anisotropic wrinkle patterns [44] for membrane thicknesses ≥ 10 nm, with characteristic length scales that depend strongly on film thicknesses. High resolution AFM topography and corresponding height profiles (Figure 2b,c) enable quantitative analysis of the wrinkle morphology across the entire thickness range, including the ultrathin 4 nm membrane. The wrinkle amplitude decreases from approximately 1800 nm for the 100 nm membrane to ≈ 50 nm for the 4 nm membrane, while the wrinkle wavelength similary decreases from ≈ 40 $\mu$m to ≈ 1 $\mu$m. This systematic evolution of wrinkle geometry with membrane thickness is well described by the buckling theory.[45, 46, 15] In contrast, when the same transfer procedure is performed using host substrates with higher elastic modulus than silicone/PET, the mechanical stress is dissipated differently, favoring crack and buckling formation rather than wrinkle development (see Figure S2).[47, 48, 49]

## 2.3 Nanomechanical properties

Then, the nanomechanical properties of the LSMO membranes supported on silicone/PET substrates were investigated by AFM-Force Distance Spectroscopy (AFM-FD). AFM-FD mapping was first performed for the 10-100 nm LSMO membranes. Figure 3a shows a representative stiffness map overlaid on the simultaneously acquired topography for a 50 nm LSMO membrane. Both wrinkle crests and valleys exhibit comparable stiffness values (K), which are systematically higher than those measured in the transition regions where distinct tip-surface contact geometry can also affect. The spatial dependence is corroborated by individual force-separation curves acquired at wrinkle crests, valleys and nominally flat regions for membranes of different thicknesses (Figure 3b).

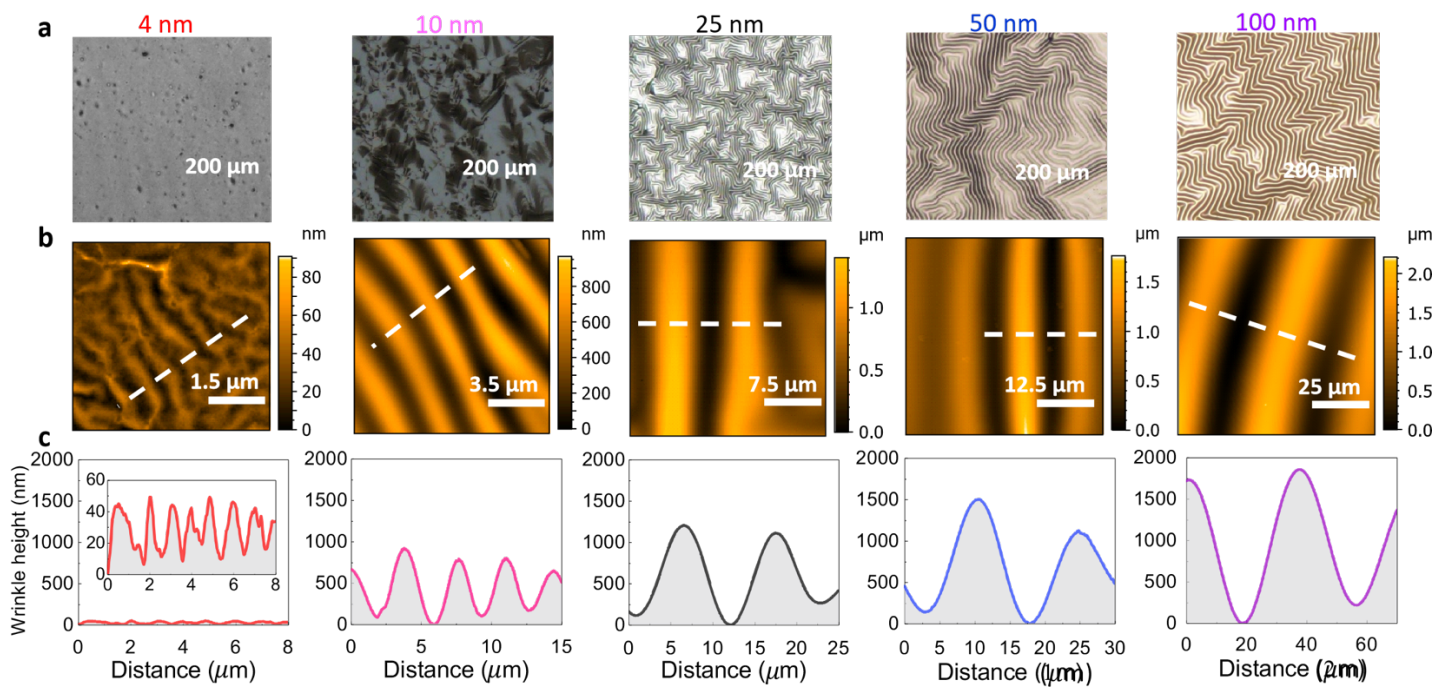

Figure 2: Surface morphology study of the 4-100 nm LSMO/PET membranes (a) Optical microscope, (b) AFM topographic images, (c) height profile from the topographic images in (b).

This trend is well captured by existing analytical models, which predict that the stiffest regions concentrate the maximum bending strain. [50, 51] On the other hand, the measured stiffness moderately increases with the membrane thickness, which is in agreement with the surface elasticity models. [52, 53] However, the measured values are substantially lower than those predicted with first-principles computational methods for bulk LSMO ($K_{LSMO}$ = 130 N/m, Methods), evidencing a pronounced effect of the compliant substrate ($K_{silicone/PET}$ < 0.5 N/m), as reported in previous nanoindentation studies performed in similar systems. [54] Repeated fracture occurred during indentation of the ultrathin 4 nm membranes, underscoring their intrinsic mechanical fragility and precluding a reliable investigation in extreme thin-limit, a regime where other freestanding membranes have exhibited anomalous nanomechanical responses. [55]

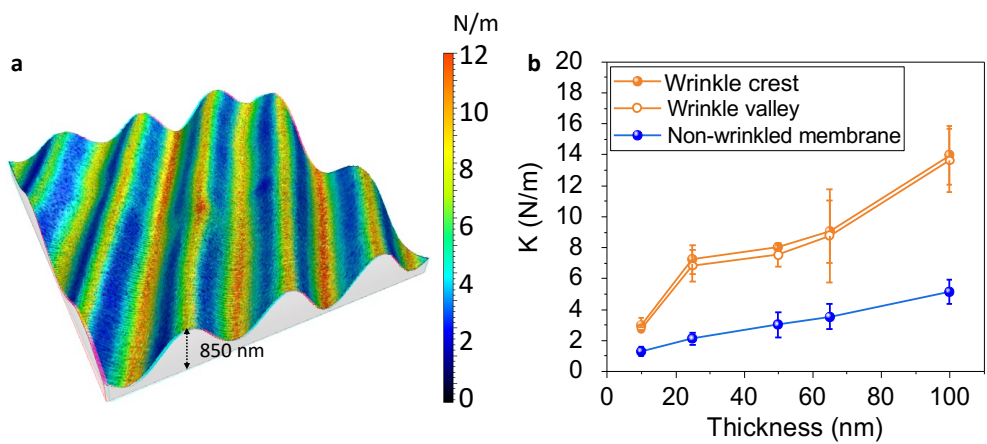

Figure 3: AFM-FD Spectroscopy of (a) 50 nm LSMO membrane on PET (b) Stiffness (K) dependence on membrane thickness (10 -100 nm).

## 2.4 Strain, atomic distortions and electronic structure

To establish a direct link between the macroscopic geometric modulation of the membranes and their physical properties, we investigated the role of local strain and atomic-scale distortions using scanning transmission electron microscopy (STEM) combined with electron energy loss spectroscopy (EELS). Cross-sectional specimens were prepared from wrinkled LSMO membranes with thicknesses of 4, 10, 15, 25 and 50 nm transferred onto Si substrates (see Methods and Figure S3, S4). Atomic-resolution high-angle annular dark-field (HAADF) STEM imaging of wrinkle crests (Figure 4a) confirms the high-crystalline quality of the films and shows distinct bending contours, indicative of variable flexural deformation. Analysis of the local in-plane ($\epsilon_{xx}$) and out-of-plane ($\epsilon_{zz}$) strain from the HAADF images (Figure 4b-c) reveals the formation of pronounced vertical strain gradients ($\epsilon_{xx,z}$ and $\epsilon_{zz,z}$). Quantification shows that both the strain and strain gradient magnitudes increase systematically as the film thickness is reduced (Figure 4e), with an abrupt enhancement below 15 nm. This trend culminates in the 4 nm films, which exhibit a maximum strain ($\epsilon_{xx}$ max) of 5.2% (See Table S3), corresponding to a $\epsilon_{xx,z}$ of $2.5 \times 10^7 \, m^{-1}$. This gradient arises from the flexure-induced variation of the lattice spacing across the films thickness, leading to a reversal of the unit cell tetragonality (c/a) between the compressed bottom (c > a) and tensile top (c < a) regions (Figure 4d). A symmetric inversion of this strain state is found in the concave valleys (Figure S5). In contrast, the shoulder regions between these features show no significant strain gradient, isolating the observed structural changes to areas of pronounced curvature, in well agreement with AFM-FD observations (Figure 3). The geometric distortion of the LSMO lattice is thus maximized in the thinnest film and gradually relaxes toward a more cubic symmetry with increasing thickness. [56]

The strain gradient in the 4 nm film is an order of magnitude larger than those typically achieved in epitaxially strained LSMO thin films [57, 58] and comparable to recent reports of giant strain gradients in ultrathin freestanding perovskites. [59] Such substantial strain gradients can couple to lattice polarization via the flexoelectric effect, inducing or enhancing polarization even in non-polar oxides. [28, 59] To probe the atomic-scale structural response, we employed annular bright field (ABF) STEM imaging, which provides sensitivity to light oxygen atoms and enables direct visualization of octahedral distortions. Figure 4g-h show ABF images of the two extreme thicknesses, 4 nm and 50 nm, along the [100] and [110] zone axes, respectively. Antiferrodistortive (AFD) octahedral rotations are a structural hallmark of metallic bulk LSMO [60, 61]. A striking common feature observed in our samples is the suppression of these rotations at the wrinkle crests and valleys (i.e. regions under high curvature strain), concurrent with the emergence of a polar distortion characterized by vertical displacements of oxygen atoms along the strain gradient direction. Conversely, AFD rotations are observed in the strain-free shoulder regions of thicker LSMO membranes (50 nm). This confirms that flexure-induced strain gradients not only suppress rotational modes but actively drives ions into a polar, non-centrosymmetric configuration.

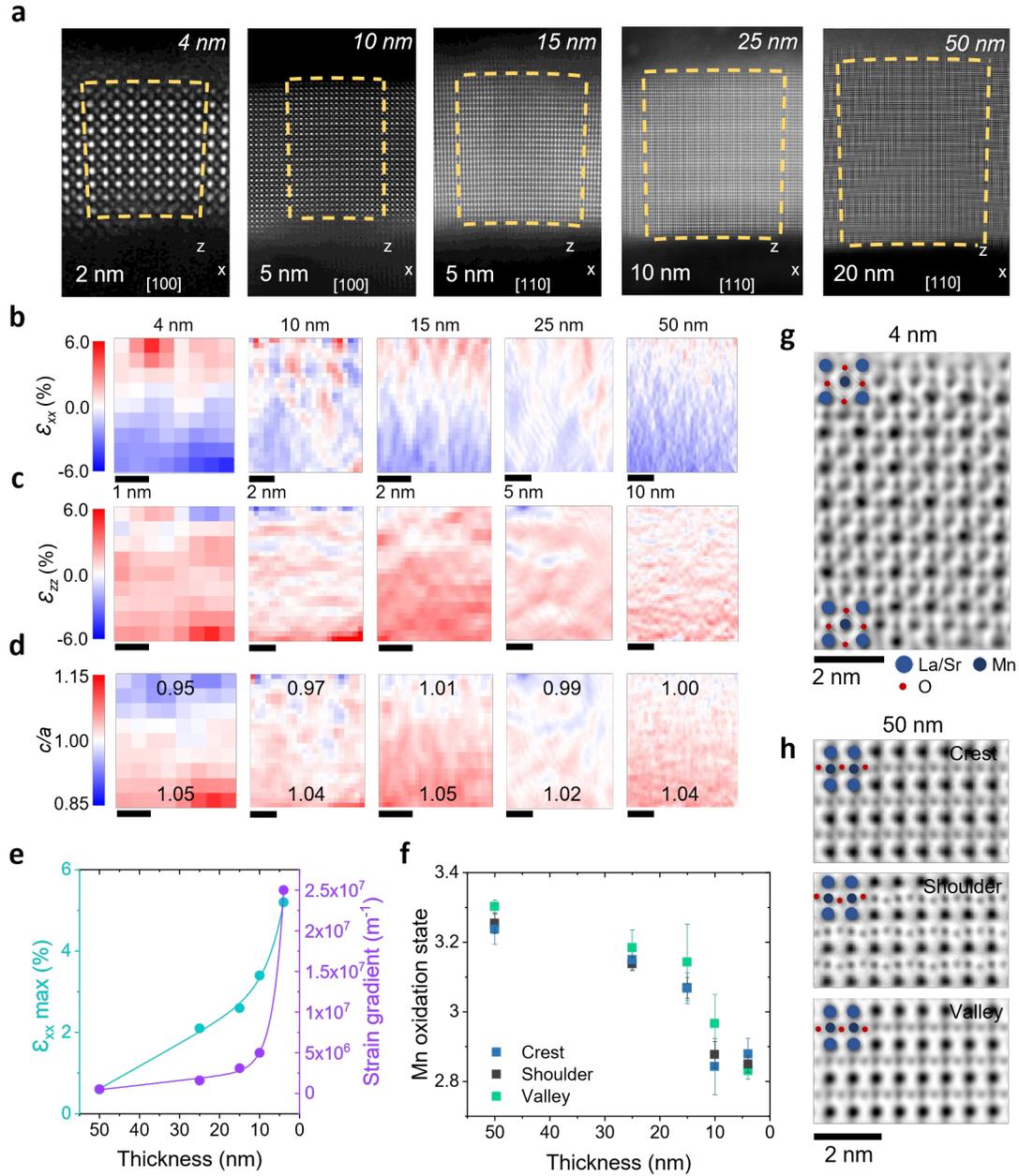

Figure 4: (a) Atomic-resolution HAADF-STEM imaging of 4, 10, 15, 25 and 50 nm membranes from the crest regions in wrinkled LSMO films. (b-d) Corresponding in plane strain (b), out-of-plane strain (c), and tetragonality (d) maps in bent regions marked with orange dashed lines in (a). (e) Maximum strain ($\varepsilon_{xx}$ max) and strain gradient as a function of film thickness. (f) Average Mn oxidation state across wrinkle topography (crest, shoulder, and valley) as a function of film thickness. (g-h) ABF images of the crest region of the 4 nm, and crest, shoulder and valley regions of the 50 nm LSMO membranes along the [100] and [110] zone axes, respectively.

In this line, electrostatic force microscopy (EFM) and Kelvin probe force microscopy (KPFM) were used to probe the local electrostatics and surface potential of the corrugated LSMO membranes, guided by their electrical behavior qualitatively assessed by conductive AFM (c-AFM). Thin membranes (4-10 nm) did not exhibit detectable electric conductivity, which prevented robust quantitative KPFM analysis due to the lack of reliable energy-level alignment. Consequently, these specimens were investigated using EFM instead. By contrast, thicker membranes (25 to 100 nm) showed measurable conductivity (details in Figure S7), thereby enabling quantitative KPFM analysis of the surface potential.

For the insulating 4 and 10 nm membranes, the topography image (Figure 5 a,e) reproduces the wrinkled morphology. The corresponding EFM phase and amplitude images reveal distinct contrast features indicating local changes in the sign and magnitude of the surface electrostatics, respectively, that spatially correlate with the wrinkle pattern, Figure 5 b,c-f,g. Control measurements performed under zero tip $V_{ac}$ bias confirm that the observed features originate from genuine electrostatic variations rather than topographic crosstalk on the EFM measurements. This correlation is further highlighted in the line profiles extracted from both topography, EFM phase and amplitude data (Figure 5 d,h) where the crests and valleys show an opposite amplitude sign and enhanced phase response compared to flat regions. This trend coincides with the existence of polar distortions identified along the strain gradient direction in the curvature analysis, in Figure 4 a-d.

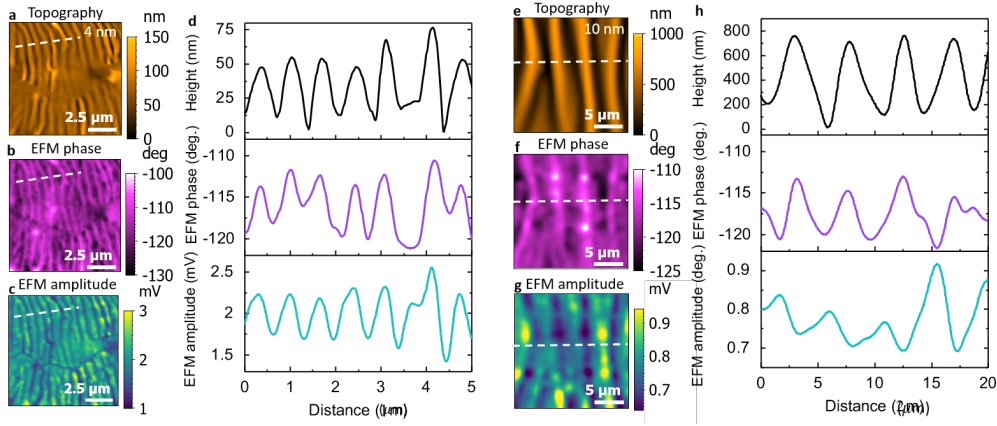

Figure 5: EFM analysis on 4 and 10 nm LSMO/PET membrane (a)topography image, (b) EFM phase image with the corresponding line profiles in (c).

Then, for thicker and electrically conductive membranes (25-100 nm), quantitative surface potential maps and the corresponding topographies were recorded by KPFM, Figure 6. The quality of the electrical grounding of the layers was additionally verified by applying an external bias and confirming the corresponding potential shift measured by KPFM. From the maps and line profiles extracted along representative wrinkles, it is revealed a correlated waved pattern between topography and surface potential, similar to that recorded by EFM for thin membranes (Figure 5). Direct comparison of EFM and KPFM analysis of 50 nm LSMO membrane can be found in Figure S8. This periodic modulation of the surface potential reinforces the formation of wrinkle-induced polar patterns arising from large strains and strain gradients. Additionally, by increasing the membrane thickness from 25 to 100 nm, the surface potential variation between crest and valley decreases from ∼ 40 mV to ∼ 15 mV, explained by a more effective screening of the internal polarizing electric field in these half-metallic LSMO membranes.[62, 63] Notoriously, these values are significantly smaller than the surface potential variations reported for insulating perovskites such as BTO and STO.[21, 30, 64, 32] These results demonstrate that wrinkled LSMO membranes can generate order polar patterns by induced strain gradients even in conductive samples.

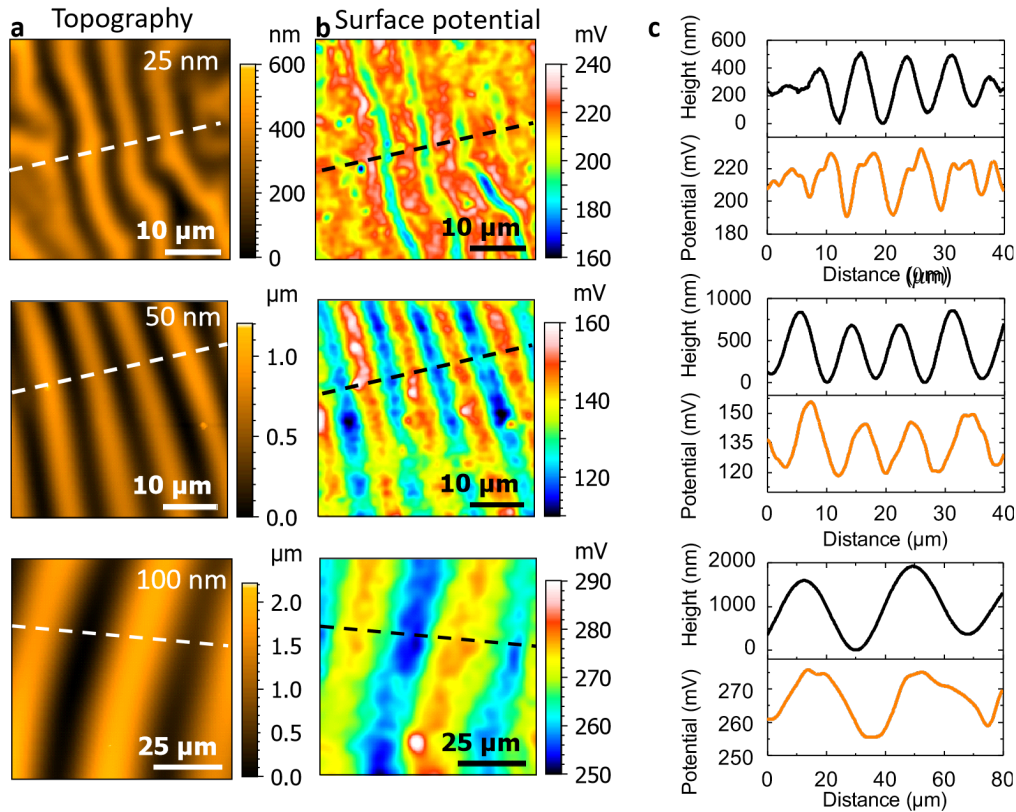

Figure 6: KPFM 25 nm, 50 and 100 nm LSMO on PET (a) topography, (b) surface potential, (c) height profile and surface potential extracted from the respective topography and surface potential maps in (a) and (b).

In view of these results, spatially-resolved electron energy-loss spectroscopy (EELS) has been employed to probe potential modifications in the electronic structure across the series of LSMO membranes.

We quantified the Mn oxidation state using the well-established L3/L2 white-line intensity ratio method (see Methods), which is sensitive to the 3d electron occupancy. Spatially resolved EELS across the wrinkle topography (crest, shoulder, and valley) reveals a profound, thickness-dependent reduction of the Mn valence (Figure 4f). The average Mn oxidation state remains near the nominal bulk value of 3.2+ for the thick 50 nm membrane but undergoes an abrupt drop to approximately 2.85+ in the thinnest membranes (4 and 10 nm). Moreover, the entire Mn-$L_{2,3}$ edge undergoes a rigid chemical shift of about 1.5 eV to lower energy loss in the 4 nm and 10 nm films compared to the thicker films (Figure S6). Both independent metrics confirm the reduction of $Mn^{4+}$ to $Mn^{3+}$ and a decisive chemical signature reminiscent of a metalinsulator phase transition. We note that these findings coincide with the thickness-dependent conductivity qualitatively identified above by scanning probe microscopy. Crucially, the Mn oxidation state is spatially uniform across the wrinkle topography for a given film thickness (Figure 4f). These results therefore suggest that the overall lowering of the electrochemical potential for oxygen vacancy formation is dominated by the film thickness rather than by variations in the local strain gradient.

Complementary evidence for hole carrier depletion is provided by the oxygen K-edge analysis Figure S6). The first pre-peak ( 530 eV), which corresponds to transitions into unoccupied O 2p states hybridized with Mn 3d orbitals, exhibits a dramatic decrease in intensity [65] for the 10 and 4 nm. This direct spectroscopic measurement corroborates the significant reduction in hole density within the Mn-O hybridized bands, consistent with the stabilization of a $Mn^{3+}$-rich state. This observation agrees with prior studies reporting non-metallic behavior in ultrathin epitaxial LSMO films, identifying the oxygen vacancy formation due to reduced dimensionality as a relevant factor. [66, 67, 68] In the present case, however, any contribution from epitaxial strain can be ruled out.

These findings reveal that thickness and wrinkle curvature are powerful levers to drive nanomechanics, electronic structure and symmetry breaking enabling to build a rich phase diagram for LSMO, see Figure 7. This scenario envisages the possibility of designing polar metals and establishing a pathway towards the flexoelectric control of electronic phases in freestanding correlated oxide membranes.[69, 70].

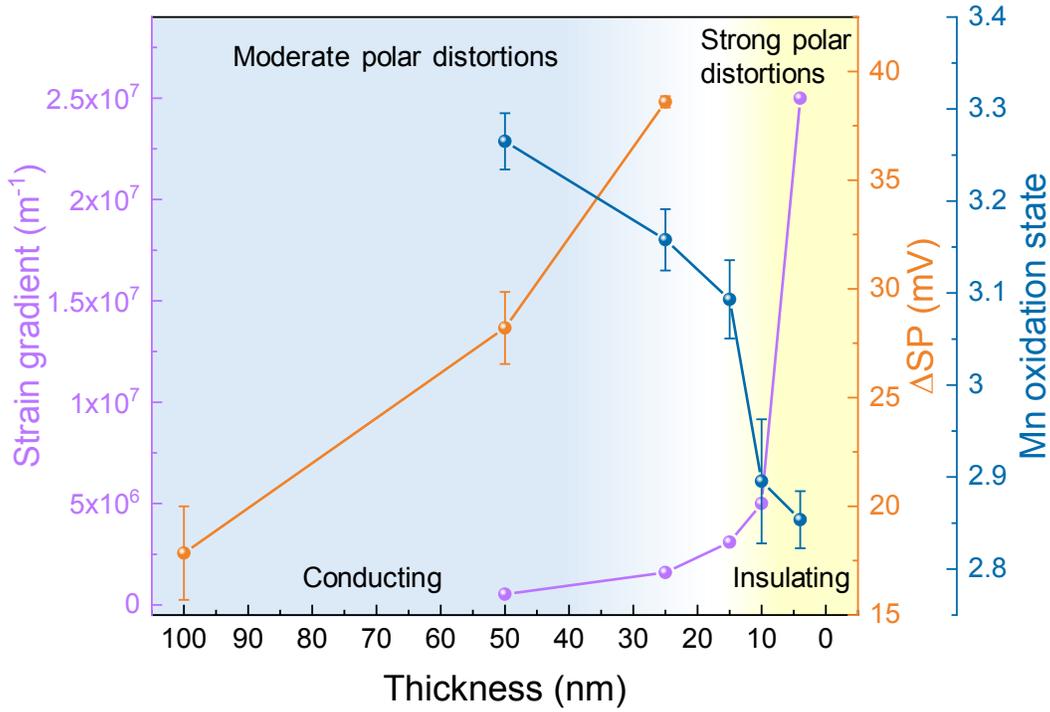

Figure 7: Thickness-dependent phase diagram of LSMO membranes spanning ultrathin film regimes (< 10 nm) characterized by extreme strain gradients, pronounced polar distortions and Mn oxidation states < 2.85+ with insulating behavior; intermediate thickness (25-20 nm) with flexure-induced strain gradients stabilizing polarity and ∼ Mn 3+ with conductive behavior ($\Delta SP_{crest-valley}$ 40-20 mV); and thickness membranes (> 50 nm) with ∼ Mn 3.2+ exhibiting conductivity ($\Delta SP_{crest-valley}$ <25 mV) alongside polar domains localized at wrinkle crests and valleys.

## 3 Conclusions

We demonstrate the fabrication of (00l) La$_{0.7}$Sr$_{0.3}$MnO$_3$ (LSMO) membranes with thicknesses ranging from 4 to 100 nm and show that their spontaneous wrinkling on compliant substrates provides a versatile route to engineer morphology-driven functionalities. The membranes preserve high crystalline quality after release and transfer, with wrinkle geometry scaling systematically with thickness. Nanomechanical mapping reveals corrugation-dependent moderate stiffness variations, with maxima at wrinkle crests and valleys, where bending strain- and the resulting strain gradients- are largest, reaching up to 5% and 2.5 x 10$^7$ m$^{-1}$, respectively, in the ultrathin limit. Atomic-resolution STEM shows that these extreme strain gradients suppress the antiferrodistortive octahedral rotations, characteristic of bulk LSMO, and induce polar lattice distortions. Scanning probe measurements further reveal that these structural distortions translate into spatially ordered electrostatic patterns that follow the wrinkle topography, demonstrating the emergence of curvature-induced polar structures even in conductive membranes. In parallel, electron energy-loss spectroscopy uncovers a thickness-dependent evolution of the electronic structure, marked with a reduction of the Mn valence state and a depletion of hole carriers in the thinnest films, fingerprint of a metal to insulator transition. Collectively, these observations define a rich phase diagram for LSMO membranes. Below 10 nm, membranes are insulating, with extreme bending strains driving pronounced polar distortions; at intermediate thicknesses (25–50 nm), membranes are conductive while flexure-induced strain gradients still stabilize polar distortions. For thicker membranes (50–100 nm), conductivity persists and polar distortions concentrate at wrinkle crests and valleys. This framework demonstrates how dimensionality and corrugations can be leveraged to cooperatively tune nanomechanics, electronic structure and symmetry breaking, establishing geometric control as a robust strategy to engineer emergent functionalities in correlated oxide systems.

# 4 Experimental Section

*Membrane preparation*:

20 nm-thick $SrCa_2Al_2O_6$ ($SC_2AO$) sacrificial layers were prepared on (001) $SrTiO_3$ (STO) single-crystal substrates by chemical solution deposition (CSD).[43] The films were subsequently transferred to a pulsed laser deposition (PLD) chamber and annealed at 825 °C for 30 min at an oxygen partial pressure $PO_2$ of 0.1 mbar. Immediately after, $La_{0.7}Sr_{0.3}MnO_3$ (LSMO) films with thicknesses ranging from 4 to 100 nm were deposited by PLD at 725 °C and a $PO_2$ of 0.1 mbar. To release the LSMO membranes, the heterostructures were attached to a silicone-coated polyethylene terephtalate (PET) sheets and immersed in Milli-Q water for 48h during which the $SC_2AO$ sacrificial layer was fully dissolved. This process enables complete transfer of the LSMO films onto the PET substrate while preserving the native wrinkled morphology.

*Membrane characterization*:

Crystal Structure: X-ray Diffraction (XRD) measurements were performed with Cu-K$\alpha$ using a Bruker AXS (model A25 D8 Discover).

Scanning probe microscopy (SPM) experiments were carried out with a Park System NX10 and FX40 in ambient conditions. All data were processed by using Mountains software.[71] AFM -topography images were acquired in non-contact and tapping mode allowing high-resolution mapping of wrinkle amplitude and wavelength. Force-Distance AFM (FD-AFM) maps were acquired over wrinkle peaks, valleys and flat regions. For each measurement, the tip approach and retraction curves were recorded and the local stiffness was extracted from the linear region (elastic regime) of the repulsive part. A single crystal sapphire surface was employed to calibrate the photodiode sensitivity factor. AppNano Forta and Nanosensors FMR tips where employed and the constant force of the cantilevers was calculated by the thermal tuning and Sader methods. Averaged values of the stiffness have been extracted from performing 20 different indentations at equivalent regions for the same thickness for the 4-100 nm range. EFM and KPFM measurements were performed by non-contact mode and single pass KPFM mode by sideband KPFM from Park Systems. Pt coated conductive PPP-NCHPt tips from Nanosensors were employed. Electrical contact to the LSMO thin film was established by mechanically pressing a soft electrode against one edge of the layer. The continuity of the film, with the absence of damage or cracks, was inspected by optical microscopy before each experiment. Proper electrical conductivity was verified by applying variable voltages to the film and measuring the response by KPFM. C-AFM experiments were performed by applying the bias voltage to the tip and measuring the resulting current at the sample using a DLPCA-200 FEMTO transimpedance amplifier. BudgetSensors ContE-G and Multi75E-G Pt-coated probes were employed to ensure accurate force control in contact mode and to prevent damage to the thin-film layers. C-AFM experiments were perfoed by applying the bias voltage to the tip and measuring the resulting current at the sample using a DLPCA-200 FEMTO transimpedance amplifier and also by applying the bias on the sample and measuring the current at the tip by an internal current amplifier of the FX40.

Sample preparation for cross-sectional STEM investigations: To preserve the wrinkled topography of the LSMO membranes for cross-sectional scanning transmission electron microscopy (STEM), we mounted them using M-Bond 610 epoxy adhesive. The epoxy was first applied to a Si substrate, after which the LSMO/PET membrane was transferred on top (see Figure S3). The epoxy was cured for 2 h at 200 ºC under moderate pressure. The PET support was then gently peeled away, leaving the wrinkle morphology intact. Cross-sectional lamellae were prepared using a standard focused ion beam (FIB) lift-out procedure in a Thermo Fisher Scientific Helios dual-beam system. Prior to milling, a 40 nm amorphous carbon layer was sputtered onto the LSMO surface, followed by protective tungsten layer: 200 nm deposited by electronbeam-assisted deposition, and then 2.5 $\mu$m deposited by Ga ion-beam-assisted deposition over the region of interest. The lamellas were transferred onto Cu grids and progressively thinned to electron transparency by gradually reducing the Ga ion beam acceleration voltage and current. This protocol maintained the native morphology of the wrinkles and yielded high-quality cross-sections suitable for electron microscopy analysis.

Aberration corrected STEM imaging and spectroscopy: Aberration-corrected STEM was performed using a Thermo Fisher Scientific Spectra 300 operated at 300 kV. High-angle annular dark-field (HAADF) and annular bright-field (ABF) images were acquired with a probe convergence semi-angle was set to 19.5 mrad. The collection semi-angles of the HAADF and ABF detectors was set to 63-200 and 10-20 mrad, respectively. To minimize scan-distortion and drift artifacts, 20-short-dwell-time image frames were recorded, then aligned and integrated by cross-correlation during post-processing. Atomic positions of the A- and B-site cations were extracted from HAADF and ABF images, respectively, using the Python library Atomap.[72] From these positions, strain maps, tetragonality and bond-angle measurements were computed. In each film, strain magnitude and tetragonality were quantified by comparing the in-plane and out-of-plane lattice parameters of wrinkled regions with those of flat, non-bent LSMO areas. While the in-plane lattice parameter varies symmetrically through the film thickness, suggesting a linear bending strain profile, the out-of-plane lattice parameter exhibits an asymmetric distribution, which would alter the relaxed lattice parameter and shift the mechanical neutral axis from the geometric mid-plane. Therefore, the in-plane strain gradient ($\varepsilon_{xx,z}$) was calculated over the total thickness using the difference between maximum tensile and compressive strains, whithout assuming the localtion of the zero-strain point.

STEM electron energy loss spectroscopy (EELS) was conducted in a JEOL NEO-ARM operated at 200 kV and equipped with a Gatan GIF continuum K3 direct-electron detection camera. Multiframe spectrum images were acquired across wrinkle topographies (peak, shoulder, and valley) under drift-correction conditions, with a dispersion of 0.09 eV/channel and an exposure time of 0.05 s per pixel (15 frames total). Signals averaged from the bottom part of the films are shown in Figure S6. Spectrum images were denoised using principal component analysis, retaining the first five principal components. Background subtraction was performed by fitting a power-law function to the pre-edge energy range. Absolute energy calibration was achieved pixel-by-pixel using the zero-loss peak (ZLP) position as reference (dual-EELS method). The ZLP was also used to correct for plural scattering effects due to sample thickness. The Mn-L3/L2 ratio was determined by fitting a Hartree-Slater cross-section double step function [65, 73]. After scaling and subtracting this function, the remaining intensities under the L3 and L2 edges were integrated over windows of 1-3 eV, and their ratio was calculated. Error bars in Figure 4f represent the standard deviation obtained by systematically varying the integration window width from 1 to 3 eV in steps of 0.2 eV. The Mn oxidation state was then derived using the exponential calibration of Tan et al. [73], which relates the L3/L2 ratio to oxidation state.

*Computational Methods*: The energetic, structural, magnetic and elastic properties of $La_{0.7}Sr_{0.3}MnO_3$ (LSMO) thin films under uniaxial strain were calculated using quantum first-principles methods based on density functional theory (DFT). The PBEsol variant of the generalised gradient approximation to DFT [74] was used as is implemented in the VASP package.[75] The "projector augmented wave" method was employed to represent the ionic cores,[76] and the following electrons were considered as valence: Mn 3d, 4s; La 5d, 5p, 6s; Sr 5s, 4p; O 2s and 2p. Wave functions were represented in a plane-wave basis truncated at 650 eV. We used a 40-atoms simulation cell in which the usual ferroelectric and anti-ferrodistortive distortions occurring in oxide compounds could be reproduced. [77, 78, 79] An on-site Hubbard correction of U=3.0 eV was employed to better describe the localised orbitals in the Mn atoms.[80] Chemical disorder was effectively reproduced by means of the virtual crystal approximation.[81] For integrations within the first Brillouin zone, we adopted a Gamma-centred k-point grid of 10x10x10. Geometry relaxations were performed by using a conjugate-gradient algorithm that optimised the volume and shape of the simulation cell as well as the atomic positions. The imposed tolerance on the atomic forces was of 0.005 eV·Å-1. By using these parameters we obtained total energies that were converged to within 0.5 meV per formula unit.

Different spin-magnetic orders were sampled on the Mn atoms, including ferromagnetic, antiferromagnetic (i.e., A, G and C type) and ferrimagnetic. Ferromagnetic ordering yielded the lowest energy for most of the simulated uniaxial strain conditions. The elastic constants of LSMO were estimated using first-principles density functional perturbation theory. [75] The mechanical stiffness of LSMO thin films, S, was estimated with the formula $S = C_{33}/h$, where $C_{33}$ represents the out-of-plane elastic modulus in Voigt notation and h the thin film thickness.

**Acknowledgements**

This work was funded through the projects Severo Ochoa CEX2023-001263-S by MICIU/AEI/10.13039/5011000 and projects PID2020-114224RB-I00, PID2023-149407NB-I00 funded by MICIU/AEI/10.13039/501100011033 and by ERDF/, UE. M.R. acknowledges the support from the FPI fellowship PRE2021-199672 funded by MICIU/AEI/10.13039/501100011033 and FSE investing in your future. The authors thank the thin film scientific service at ICN2. Authors acknowledge the use of instrumentation as well as the technical advice provided by the Joint Electron Microscopy Center at ALBA (JEMCA) and funding from Grant IU16-014206 (METCAM-FIB) to ICN2 funded by the European Union through the European Regional Development Fund (ERDF), with the support of the Ministry of Research and Universities, Generalitat de Catalunya. C.C. acknowledges support by MICIN/AEI/10.13039/501100011033, under the grants PID2023-146623NB-I00, PID2023-147469NB-C21 and the Maria de Maeztu Units of Excellence Programme CEX2023-001300-M, and by the Generalitat de Catalunya, under the grants 2021SGR-00343, 2021SGR01519 and 2021SGR-01411. Computational support was provided by the Red Española de Supercomputación under the grants FI-2025-1-0015, FI-2025-2-0006, FI-2025-2-0028 and FI-2025-3-0004. This work was supported by grants awarded through the ALBA (20250340089; 20250350036), and Integrated Infrastructure for Electron Microscopy of Materials, ELECMI (ELC447-2024; ELC561-2025), open calls for proposals. This work has been done in the framework of the doctorate in material science of the Autonomous University of Barcelona (M.R.). Chat GPT was used to support text editing and proofreading. All content was reviewed and approved by the authors.